\documentclass[9pt,cites,graphicx]{article}

\usepackage{epsfig}

\title{Coarse Grained Modeling of The Interface Between Water and Heterogeneous Surfaces}

\author{Adam P. Willard and David Chandler \\[3mm]Department of Chemistry, University of California, Berkeley, California 94720 \\[1mm]}

\begin{document}

\newcommand{\kB}{k_{\mathrm{B}}}
\newcommand{\me}{\mathrm{e}}
\newcommand{\K}{\mathrm{K}}
\newcommand{\diff}{\mathrm{d}}
\newcommand{\mhxy}{\bar{h}_{\ba}^{(\alpha)}}
\newcommand{\ba}{\mathbf{a}}
\newcommand{\br}{\mathbf{r}}

\maketitle

\renewcommand{\thefootnote}{\fnsymbol{footnote}}

\noindent 
Using coarse grained models we investigate the behavior of water adjacent to an extended hydrophobic surface peppered with various fractions of hydrophilic patches of different sizes.  We study the spatial dependence of the mean interface height, the solvent density fluctuations related to drying the patchy substrate, and the spatial dependence of interfacial fluctuations.  We find that adding small uniform attractive interactions between the substrate and solvent cause the mean position of the interface to be very close to the substrate.  Nevertheless, the interfacial fluctuations are large and spatially heterogeneous in response to the underlying patchy substrate.  We discuss the implications of these findings to the assembly of heterogeneous surfaces.

\section{\label{sec:introduction}Introduction:  Interfaces and Hydrophobicity}

The interaction of liquid water with oily components in aqueous solution is
central to many phenomena~\cite{tanford}. These phenomena are called
\textquotedblleft hydrophobic effects." The hydrophobic effects leading to
robust assembly, such as micelle formation and protein aggregation, follow
from the nucleation of water-vapor-like interfaces adjacent to sufficiently
extended hydrophobic surfaces~\cite{DC05}. This article reports on
computer simulation studies of the behavior of such interfaces when the
surfaces contain hydrophilic regions.

Our analysis focuses on probability distributions for density fluctuations
in water, distributions like $P_{v}(N)$, which stands for the probability that the centers of 
$N\,$water molecules are found in a volume $v$. We consider such distributions because arrangements
of water molecules near hydrophobic solutes are similar to those of
water near voids in the liquid.  The connection between voids and hydrophobicity is found in Stillinger's~\cite{FS73} proposal that a water-oil interface is similar to a water-vapor interface.  This proposal is supported by recent experiments~\cite{granick1,netz1,MM06}.

Hummer, Pratt and their coworkers~\cite{GH96} were the first to
call attention to the behavior of $P_{v}(N)$. For small microscopic
volumes in water, they used computer simulation to show that this probability
is almost exactly Gaussian. Some theories of homogeneous liquids and
hydrophobic effects are based upon the assumption that microscopic
density fluctuations obey Gaussian statistics~\cite{DC93}, so the finding
of Ref.~\cite{GH96} provides support for those theories. But more important is a
remarkably simple yet quantitatively accurate theory for solvation free
energies of small hydrophobic species~\cite{GH96,SG96}. This consequence follows from the fact that the solvation free energy or excess
chemical potential for a solvent excluding volume $v$, $\Delta \mu _{v}$, is
given by~\cite{BW63}
\[
\beta \,\Delta \mu _{v}=-\ln P_{v}(0),
\]%
where $1/\beta =k_{{\rm B}}T$ is Boltzmann's constant times temperature.
Therefore, to the extent that $P_{v}(N)$ is Gaussian, $\Delta \mu _{v}$ can
be expressed entirely in terms of bulk water's mean density and mean-square
density fluctuations, both of which can be determined from experimentally
known quantities. With this solvation free energy in hand, hydration free
energies of real apolar molecules can be estimated, using perturbation
theory to account for solvent-solute forces beyond those of excluded volume
interactions~\cite{DC05,LP77}.

But for large volumes $v$, or for small volumes in proximity to an
extended hydrophobic surface, the Gaussian approximation to the probability $%
P_{v}(N)$ ceases to be accurate. Here, probabilities for fluctuations are
not the same as those for the bulk liquid because the chemical potentials of liquid water and vapor differ by very little compared to $k_{{\rm B}}T$. As such, a
large void in water can nucleate vapor-like configurations~\cite{FS73},
and vapor-like configurations increase the likelihood of large density
fluctuations. Thus, for large $v$ and $N$ much smaller than its mean, $\left\langle
N\right\rangle _{v}$, $P_{v}(N)$ is much larger than the probability predicted by the
Gaussian approximation. In other words, for large enough $v$, $P_{v}(N)$
possesses a fat non-Gaussian tail at small values of $N$. Understanding the
nature of this tail is important for estimating values of solvation free
energies of sufficiently large clusters and extended surfaces of hydrophobic
species. It is also important for estimating likely pathways by which water
molecules are displaced during assembly of hydrophobic clusters.

A theory for the fat tail or large length scale hydrophobicity must account
for the presence of both liquid and vapor and the interface between them.
The theory of Lum, Chandler and Weeks (LCW)~\cite{LCW} does so by partitioning
the density field of water into two components. One is the field $n_{{\rm s}%
}({\bf r})$, which describes two-phase coexistence and is presumed to vary
slowly in space (hence the subscript \textquotedblleft s"). The other, $%
\delta \rho ({\bf r})$, is a Gaussian field, which is assumed to include all
variation not captured by $n_{{\rm s}}({\bf r})$. The theory predicts the
onset of large length scale hydrophobicity at a size of roughly 1 nm.
Variations of $n_{{\rm s}}({\bf r})$ occur on length scales greater than 0.2
or 0.3~nm. The equations governing the behavior of $n_{{\rm s}}({\bf r})$ are
continuum versions of equations governing a binary field (i.e., the density field of a lattice gas or the equivalent spin field of an Ising model) on a three dimensional cubic lattice~\cite{IMSM,CandL}. To the extent that a spatially coarse
resolution of the liquid density is of interest, a lattice gas field suffices to
describe the fluctuations of water. The role of $\delta \rho ({\bf r})$ in
that case is implicit, affecting the values of parameters of the lattice gas
Hamiltonian. We have adopted this approach in earlier work modeling dynamics
of a hydrophobic chain in water~\cite{PRtW02}, dynamics of water in and around
nanometer scale tubes~\cite{LM04}, and the dynamics of dimerization for two
nanometer scale spheres in water~\cite{AW08}. Some additional justification
for the approach can be found in theoretical analysis~\cite{tenwolde1}
and explicit atomistic modeling~\cite{TFM07}. Examples of other applications of the
lattice gas model to treat water fluctuations near extended hydrophobic
surfaces are found in Refs.~\cite{AL00}.

The models we employ are defined in the next section. We then present results of this general coarse grained approach, examining density
fluctuations adjacent to extended complex surfaces in a way that elucidates recent atomistic modeling~\cite{NG08,NG07}. Some of our calculations focus
directly on the fluid interface, i.e., a $d=2$ dimensional manifold in $d=3$
dimensional space. Our principal results demonstrate the variability of
probabilities for dewetting fluctuations, and further show that the behaviors of
these probabilities are richer and more physically pertinent than might be
guessed from the behaviors of their mean values alone.

\section{\label{sec:model}Models}
\subsection{Lattice gas}
As noted, we use the lattice gas model as the discrete version of the slowly varying field of the LCW theory~\cite{LCW}, with the parameterization of Ref.~\cite{tenwolde1}.  In particular the density field is taken as a binary field on a three dimensional cubic lattice with lattice spacing $\l$.  Positions of a lattice site are specified with $\br= x \hat{\mathbf{x}} + y \hat{\mathbf{y}} + z \hat{\mathbf{z}}$ where $x$, $y$, and $z$ are Cartesian components.  In units of $\l$ these components have integer values.  The lattice site at position $\br$ has occupation number $n_\br$ equal to 1 if the volume within that site is liquid-like and equal to 0 if the volume within that site is vapor-like. This lattice gas model is capable of supporting liquid-vapor phase coexistence  and its interfaces.  The energy for a given binary density field $\lbrace n_\br \rbrace$ is given by, 
\begin{equation}
E_{\mathrm{L}}(\lbrace n_\br \rbrace) = -\epsilon {\sum_{\br,\br'}}' n_\br n_{\br'} - \sum_\br \mu n_\br,
\end{equation}
where the primed summation is taken over nearest neighbor pairs of lattice sites, and at coexistence the chemical potential, $\mu$, has value $3 \epsilon$.  To match the surface tension, compressibility and proximity of water to coexistence, we use $\epsilon = 1.51 \kB T$, $\l = 0.21$~nm and $\mu = 3 \epsilon + 2.25 \times 10^{-4} \kB T$.  In the simulation of this model, we use cells that are periodically replicated in the $x$ and $y$ directions, while the upper $z$ boundaries are populated with liquid-like lattice sites and the lower $z$ boundaries serve to characterize patchy surfaces described below.  

  Fluctuations across a liquid-vapor interface can be correlated over very long distances while correlation lengths in the bulk liquid are very small.  Thus, fluctuations of the interface can be interesting even while those in the bulk are uninteresting.  For this reason we choose to simulate systems with large interfacial area (in the $xy-$plane) and relatively little bulk liquid.  Taking the perspective to the extreme would disregard bulk entirely, as is done with an interface model, which we turn to now.

\subsection{Fluid Interface}
Here we adapt the two-dimensional model of a liquid-vapor interface described by Weeks~\cite{weeks1}.  In this model, a liquid-vapor interface is treated as a two-dimensional manifold that is coarse grained onto a square lattice in a reference plane with lattice spacing approximately equal to the bulk correlation length of water, $\xi = 0.42 \mathrm{nm}$.  This two-dimensional interface is periodically replicated in each of the two dimensions of the reference $xy-$plane.  The interfacial profile is characterized by the set of height variables $\lbrace h_\ba \rbrace$, where $h_\ba$ is the distance of the interface normal to the reference $xy-$plane, and $\ba = x \hat{\mathbf{x}} + y \hat{\mathbf{y}}$ refers to a lattice point on that plane.  In this model, which we call the ``fluid interface model,'' the free interface has the energy function,
\begin{equation}
	E_{\mathrm{W}}(\lbrace h_\ba \rbrace) = \frac{\Gamma}{2} {\sum_{\ba,\ba'}}' (h_\ba - h_{\ba'})^2,
\end{equation}
where the primed summation is taken over nearest neighbor pairs, and to be consistent with the surface tension of water, we use $\Gamma \xi^2 = 0.1 \kB T$.

\subsection{Patchy substrate}
We couple the models described above to a patchy substrate.  The substrate interacts with the solvent through excluded volume interactions as well as attractive potentials.  The substrate lies in the $z=0$ plane, and is taken to be effectively infinite in the $xy-$plane, and it prevents the liquid (and its interface) from existing at $z<0$. In addition to excluding volume, the substrate surface contains regions which attract the liquid or its interface.  These are called ``hydrophilic'' regions.  The other portions of the substrate, which simply exclude the liquid, are the ``hydrophobic'' regions.  The surface is partitioned with the same length scale as the solvent and the state (hydrophilic or hydrophobic) is specified with a variable $\sigma_\ba$ (1 or 0, respectively).  For the lattice gas model, the interaction energy between the patchy substrate $\lbrace \sigma_\ba \rbrace$ and the first (i.e., $z=1$) layer of lattice sites is, 
\begin{equation}
\Delta E_\mathrm{L}(\lbrace n_\br \rbrace; \lbrace \sigma_\ba \rbrace) = -\sum_\ba (\epsilon \sigma_\ba n_{\ba,1} + \Delta \gamma n_{\ba,1}).
\label{eq:latatt}
\end{equation}
Here, $n_{\ba,z}$ refers to $n_\br$ for $\br = \ba + z \hat{\mathbf{z}}$, and $\ba$ specifies $x$ and $y$.  The strong adhesive forces used to mimic the hydrophilic interactions come from the first term of Eq. \ref{eq:latatt}.  The quantity $\Delta \gamma$ is the surface adhesive interaction which is included to capture small attractive interactions such as van der Waals attractions between oil and water.  In the absence of hydrophilic regions ($\sigma_\ba=0$ for all $\ba$), $\Delta \gamma \langle n_{\ba,1} \rangle / \l^2$ is the mean field estimate of the difference between water-oil surface tension and water-vapor surface tension.  In this case (all $\sigma_\ba$ set to zero), the choice $\Delta \gamma \approx 0.05 \epsilon$, gives an average density profile $\langle n_{\ba,z} \rangle$ consistent with that found experimentally for water-oil interfaces\cite{granick1}.

For the fluid interface model, the interaction energy between the patchy substrate $\lbrace \sigma_\ba \rbrace$ and the interfacial profile is,
\begin{equation}
	 \Delta E_{\mathrm{W}}(\lbrace h_\ba \rbrace; \lbrace \sigma_\ba \rbrace) = W(\lbrace h_\ba \rbrace) - \sum_\ba \Theta(h_\mathrm{c} - h_\ba)(\sigma_\ba \mathcal{E} \xi^2 + \Delta \Gamma \xi^2) ,
\end{equation}
where $\Theta (x)$ is the Heaviside step function which is equal to 1 for $x > 0$ and zero otherwise, $h_\mathrm{c}$ is a cutoff distance equal to the bulk correlation length, the hydrophilic interaction energy $\mathcal{E} \xi^2$ is taken to be $3 \kB T$ (which amounts to about half the attractive energy per surface unit of $\epsilon$ for the lattice gas), and the function $W(\lbrace h_\ba \rbrace)$ exludes the interface from crossing the $z=0$ plane (it is infinite if any $h_\ba < 0$, and it is zero if all $h_\ba \ge 0$).  The surface adhesion $\Delta \Gamma$ is included to capture small attractive interactions (as above).  For the case of no hydrophilic patches, the choice of $\Delta \Gamma \xi^2 = 0.7 \kB T$ gives an interfacial profile which is consistent with experimental observations for water adjacent to oil~\cite{granick1}.

\subsection{Monte Carlo}
For the lattice gas model, we carry out Monte Carlo trajectories for $\lbrace n_\br \rbrace$.  Acceptance rejection obeys detailed balance for the grand canonical ensemble with net energy function,
\begin{equation}
	E(\lbrace n_\br \rbrace, \lbrace \sigma_\ba \rbrace) = E_\mathrm{L}(\lbrace n_\br \rbrace ) + \Delta E_\mathrm{L}(\lbrace n_\br \rbrace; \lbrace \sigma_\ba \rbrace).
\end{equation}
To specify hydrophilic patches, i.e., to specify $\lbrace \sigma_\ba \rbrace$, we tile the substrate with $d \times d$ squares thus creating a square lattice with lattice spacing $d$, where $d$ is an integer multiple of $l$.  At random, a fraction $f$ of the $d \times d$ squares are made hydrophilic (i.e., $\sigma_\ba = 1$ for each substrate lattice site in the square).  The pattern so formed is then held fixed while the Monte Carlo trajectory is carried out for all $n_\br$'s.  Averages over these trajectories are then recorded, results of which are described in the next section.  

For the fluid interface model, we carry out Monte Carlo trajectories for $\lbrace h_\ba \rbrace $.  Acceptance and rejection obeys detailed balance with the net energy function,
\begin{equation}
E(\lbrace h_\ba \rbrace, \lbrace \sigma_\ba \rbrace) = E_\mathrm{W}(\lbrace h_\ba \rbrace) + \Delta E_\mathrm{W} ( \lbrace h_\ba \rbrace; \lbrace \sigma_\ba \rbrace).
\end{equation}
Here, the hydrophilic patterns are created as above, but now with underlying lattice spacing $\xi$ rather than $\l$.  Monte Carlo trajectories for $h_\ba$'s are performed and analyzed with $\lbrace \sigma_\ba \rbrace$ fixed.  The $h_\ba$'s evolve continuously, unlike the $n_\br$'s, which change discontinuously between 0 and 1.

\section{Mean interfacial height for various substrates}
We have chosen to focus on surfaces which can be characterized by two parameters.  These parameters are the overall fraction $f$ of hydrophilic sites and the size $d$ of the hydrophilic patches.  The interface height of the lattice gas, $h_\ba$, is defined such that $h_\ba + \l$ is the smallest value of $z$, for $z > 0$, where $n_{\ba,z}$ is not zero.  For example, if only the cell immediately adjacent to the surface at $\ba$ is empty, $h_\ba = \l$.  If none were empty, $h_\ba = 0$.  In other words, $h_\ba$ is the value of $z$ for the occupied lattice facet closest to the patchy surface at $\ba=(x,y)$.  Consider first the average interfacial height $\langle h \rangle_{f,d}$, where $\langle \cdots \rangle_{f,d}$ denotes the equilibrium average with patchy surfaces characterized by $f$ and $d$.  Specifically, 
\begin{equation}
	\langle h \rangle_{f,d} = \frac{1}{N_\mathrm{rep}} \sum_{\alpha=1}^{N_\mathrm{rep}} \frac{1}{N_\mathrm{surf}} \sum_\ba \mhxy,
\end{equation}
where the first summation is over $N_\mathrm{rep}$ different realizations of the surface patterns, $\alpha$ refers to a specific realization, i.e., $\alpha = \lbrace \sigma_\ba \rbrace$, consistent with $f$ and $d$, and the second summation is over the $N_\mathrm{surf}$ lattice sites in the patchy surface ($xy-$plane).  The quantity $\mhxy$ is the mean height of the interface over the surface site $\ba=(x,y)$, for a specific realization,
\begin{equation}
\mhxy= \frac{1}{N_\mathrm{obs}} \sum_{\tau=1}^{N_\mathrm{obs}} h_\ba^{(\alpha)}(\tau),
\end{equation}
where the summation here is over the $N_\mathrm{obs}$ averaging time steps in a single Monte Carlo trajectory.

\begin{figure}[ht]
  \begin{center}
  \includegraphics{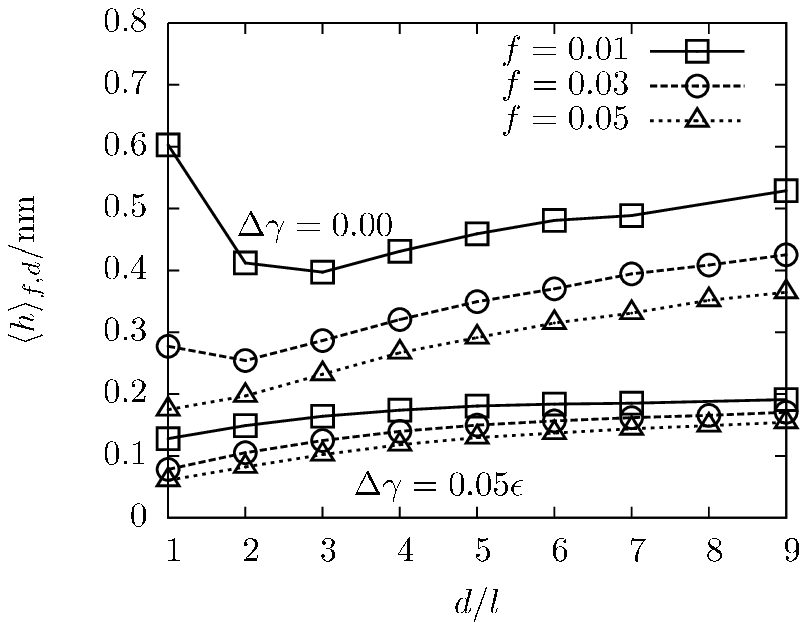}
  \includegraphics{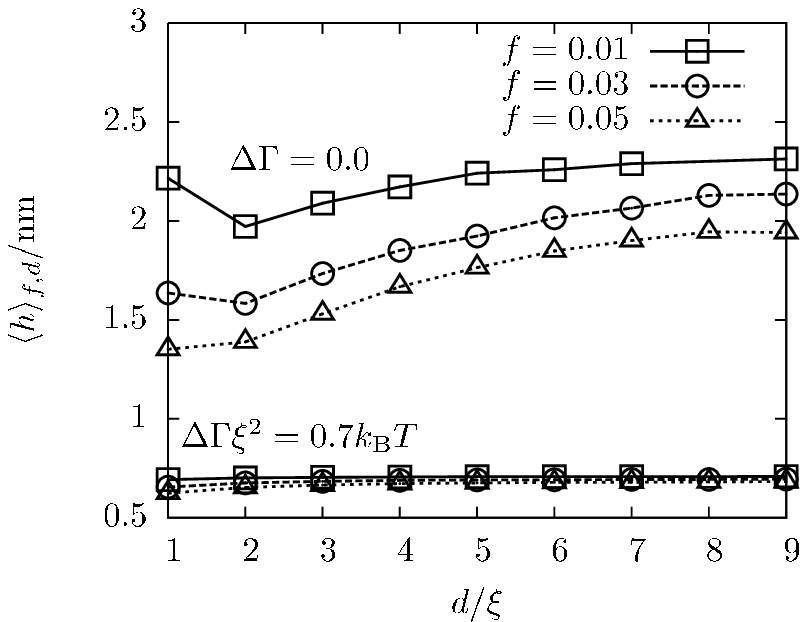}
  \caption{The equilibrium height, $\langle h \rangle_{f,d}$, of the liquid vapor interface over the surface layer as a function of patch size $d$ for the lattice gas model (top) and the fluid interface model (bottom) at different hydrophilic fractions $f$.  In each panel, the top three of curves are computed in the absence of weak water-substrate adhesive interactions, i.e., $\Delta \gamma = \Delta \Gamma = 0$, and the bottom three curves are computed with weak water-substrate adhesive interactions.}
  \label{fig:meanheight3}	
  \end{center}
\end{figure}

In the absence of weak surface adhesive interactions, i.e., $\Delta \gamma = \Delta \Gamma = 0$, for some fraction of hydrophilic coverages, $f$, the mean interface height $\langle h \rangle_{f,d}$ is non-monotonic in the hydrophilic patch size $d$.  The results for these cases are shown in Fig. \ref{fig:meanheight3}.    This non-monotonic behavior, which is observed in both the fluid interface model as well as the lattice gas model, demonstrates that for fixed surface composition $f$ there is an optimal patch size $d$ for attracting the fluid.

\begin{figure}[ht]
  \centering
  \includegraphics{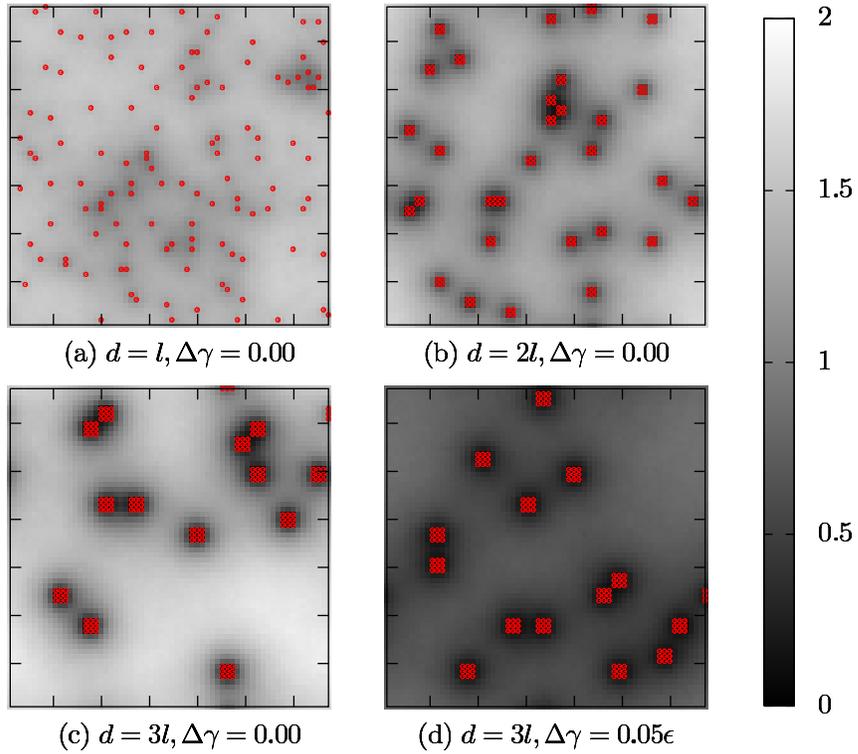}
  \caption{Mean interfacial height, $\mhxy$ projected onto the $xy-$plane for hydrophilic fraction $f = 0.03$ and $d = \l$ (top left), $d=2 \l$ (top right), and $d=3 \l$ (bottom left and right).  The height $h$ is indicated by the shading (scale at right)and the location of hydrophilic surface sites (substrate sites with $\sigma_\ba = 1$) are shown with red circles.  Panels (a), (b), and (c) correspond to averages in the absence of surface adhesive interactions  ($\Delta \gamma = 0.0$) and panel (d) corresponds to an average with surface adhesive interactions ($\Delta \gamma = 0.05 \epsilon$).  The surfaces pictured are $64\times64 \; \l^2$ in size.}
  \label{fig:height1}
\end{figure}

The mean interface height $\langle h \rangle_{f,d}$ is an average over a surface which is locally heterogeneous.  The influence of the local surface structure on the mean $\langle h \rangle_{f,d}$ can be visualized for a surface realization $\alpha$ with the spatially resolved average interface height $\mhxy$.  Figure~\ref{fig:height1} shows $\mhxy$ for representative patchy surfaces with $f = 0.03$ and $d=\l, 2\l, $ and $3\l$.  The figure shows that a hydrophilic patch affects interface fluctuations beyond the region immediately above the patch.  In Ref.~\cite{NG07}, atomistic simulations were used to study water confined between nano-scale heterogeneous surfaces.  Among their findings are that the solvent density within the first hydration layer over a nano-scale hydrophobic region is increased significantly by the introduction of a border of hydrophilic sites.  This effect is not symmetric with respect to exchange of hydrophilic and hydrophobic material as the hydration layer over a nano-scale hydrophilic region is affected very little by the introduction of a hydrophobic border.  We have found (not shown here) that these results are duplicated with the lattice models used here.

The non-monotonic behavior seen in Fig.~\ref{fig:meanheight3} can be understood qualitatively through the mean interfacial profile $\mhxy$ (shown in Fig. 2).  For a surface with $f=0.03$ and $d=l$ in the absence of surface adhesive interactions the mean interfacial profile is only slightly perturbed by the underlying hydrophilic patches.  Locally, regions of the surface with a relatively high density of hydrophilic patches lower the average height of the interface.  There are however, many isolated hydrophilic patches for which the mean interface height shows little response.  For $d = 2\l$, however, the hydrophilic patches pin the local interface, and because $d$ is relatively small there are many such patches so that the distance between a patch and its neighbors is not very large.  At $d = 3\l$ there are fewer hydrophilic patches to pin the interface and thus the hydrophobic domains are generally larger than when $d = 2\l$.
The effect of adding a small attractive interaction between the surface and solute is significant as seen in Figs.~\ref{fig:meanheight3} and \ref{fig:height1}.  The mean interface is pulled much closer to the patchy surface when weak attractions are present.  This result is consistent with the role of weak attractions uncovered in earlier work~\cite{DC05,LM07,DH02,AP01}.  The non-monotonic behavior seen in the average interfacial height is eliminated and the fluctuations are reduced.

\section{Density fluctuations and dewetting}
To explore the onset of drying-like phenomena we have considered a finite patchy substrate.  The specific substrate size is $32 \times 32 \;\l^2$ for which we have studied density fluctuations in the adjacent lattice gas.  In particular, we have computed the probability distribution, $\mathcal{P}_\alpha(\rho)$ for various substrate realizations, $\alpha$, corresponding to different hydrophilic fractions $f$ and patch sizes $d$.  The variable $\rho$ refers to the density in the first two solvent layers above the substrate,
\begin{equation}
\mathcal{P}_\alpha(\rho) =  \langle \delta ( \rho - \frac{1}{2 N_\mathrm{S}} \sum_\ba [n_{\ba,1} + n_{\ba,2}] )  \rangle_\alpha,
\end{equation}
where $\delta(\cdots)$ stands for Dirac's delta function, $N_\mathrm{S}$ is the number of lattice sites in a single layer of the lattice gas (in this case $N_\mathrm{S} = 32^2$), and the averaging implied by $\langle \cdots \rangle_\alpha$ is carried out by umbrella sampling~\cite{FandS} with the surface realization $\alpha$ fixed.  This distribution coincides with $P_v(N)$ discussed in the Introduction, in this case where $v$ is the combined volume of the first and second layers of cells above the substrate with pattern $\alpha$, and $\rho$ times that volume is $N$.

\subsection{Without weak surface adhesive interactions, $\Delta \gamma = 0$}
The inset to Fig.~\ref{fig:PofN} shows $\mathcal{P}_\alpha(\rho)$ for the case of $\Delta \gamma = 0$ (plotted with open symbols) for representative surfaces with $f = 0.03$ and $d = \l$, $3\l$, and $5\l$.  These distributions are very broad, showing significant fluctuations over a large range of densities.  These broad distributions arises because the liquid-vapor interface wanders in and out of the observed volume.  The distribution with $d=\l$ shows an additional narrow peak near $\rho = 0$, which corresponds to the dewetted state where the interface has completely pulled away from the patchy surface.

The bimodal character of $\mathcal{P}_\alpha(\rho)$ is related to the non-monotonic behavior seen in Fig.~\ref{fig:meanheight3}.  That is, the metastable ``dry'' state can emerge when the hydrophilic patch sizes are small, but this state is absent in the case of larger patch sizes.  The loss of the metastability occurs when hydrophilic patch sizes are large enough to pin the adjacent interface.  Consider the effect of excluding volume from a fluctuating liquid-vapor interface, as the substrate does.  This imposes constraints on the configurations accessible to the interface, and the closer the interface is to substrate the more severe the constraints.  Thus, for a purely hydrophobic substrate there is an entropic force that drives the liquid-vapor interface away from the surface, thereby drying the substrate.  By adding hydrophilic patches, the entropic driving force is overcome by energetically favorable interactions between the interface and the substrate, but just barely for the case of $f = 0.03$ and $d =\l$, as is evident in Fig.~\ref{fig:PofN}.

\subsection{With weak surface adhesive interactions}
The inset of Fig.~\ref{fig:PofN} also shows $\mathcal{P}_\alpha(\rho)$ for the same surface parameters, $f = 0.03$ and $d=\l$, $3\l$, and $5\l$, but with the interface adhesion turned on, i.e., $\Delta \gamma = 0.05 \epsilon$ (plotted with filled symbols).  With the interface pulled close to the surface the distributions shift to larger values of $\rho$ (more liquid-like).  For $d=\l$ the addition of adhesive interactions destabilizes the ``dewetted" state and as a result the distribution is unimodal and considerably more narrow than when $\Delta \gamma = 0$.  The main panel in Fig.~\ref{fig:PofN} shows $\log[\mathcal{P}_\alpha(\rho)]$ for $f = 0.03$, and $d=\l,3\l$, and $5\l$.  The distributions exhibit non-Gaussian tails which are a characteristic of density distributions near phase coexistence~\cite{holdsworth1,DH00}.  The departure from Gaussian behavior arises because density fluctuations in the presence of a liquid-vapor interface occur through the translation of the interface.  Fluctuations without the interface require compressing a nearly incompressible fluid.  The non-Gaussian tails imply a far greater probability for large density fluctuations near the substrate than for those in the bulk.

The quantity $\log[\mathcal{P}_\alpha(\rho)]$ is proportional to the free energy to alter the solvent density in the volume adjacent to the substrate, and thus reflects the cost to dry the patchy substrate.  There is a trivial dependence of the free energy difference between the $\rho = 0$ and $\rho = 1$ state on the hydrophilic fraction $f$.  For fixed $f$, however, the shapes of the distributions depend on the hydrophilic patch size $d$, and this feature is pertinent to the kinetic pathways to drying.  Moving from the wet state to the dry state as a function of $\rho$, initially near $\rho \approx 0.9$, a surface with larger $d$ incurs less of a free energy cost to dry than for a surface with a relatively small $d$.  As $\rho$ approaches $\rho = 0$, however, for the system with smaller $d$, fluctuations to lower $\rho$ have lower free energy than for systems with larger $d$.  As the solvent density, $\rho$, over the patchy substrate is reduced, the substrate and the hydrophilic patches dry.  The solvent density over the hydrophilic patches, the patches where $\sigma_\ba = 1$, is given by
\begin{equation}
m_1 = \frac{1}{f N_\mathrm{S}} \sum_\ba \sigma_\ba n_{\ba,1}.
\end{equation}
Figure \ref{fig:rho_dist} shows the average value of $m_1$ for substrate realization $\alpha$, $\langle m_1 \rangle_{\rho,\alpha} $, as a function of $\rho$, where the average implied by $\langle \cdots \rangle_{\rho,\alpha}$ is an average over configurations with fixed $\rho$, and substrate realization $\alpha$.  In Fig. \ref{fig:rho_dist} we see that when $d = \l$, the hydrophilic patches dry essentially uniformally as the solvent density, $\rho$, is decreased.  When the hydrophilic patches are large, however, $\langle m_1 \rangle_{\rho,\alpha}$ does not show a significant response to $\rho$ until $\rho$ is quite small, indicating that when $d$ is large the hydrophilic patches are among the last regions to dry.  Qualitatively, therefore, the initial stages of drying are easier for systems with larger $d$ because there are relatively large hydrophobic domains from which to pull the interface.  The final stages of drying are difficult when large hydrophilic patches exist, and comparatively easy for a surface with poorly pinning small patches.

\section{Spatial dependence of height fluctuations}
The distribution, $\mathcal{P}_\alpha(h;\ba)$ is the probability for observing interface height $h$ over surface site $\ba=(x,y)$ for surface pattern $\alpha$.  To visualize these distributions we define the free energy $A_\alpha(h;\ba)$, 
\begin{equation}
	A_\alpha(h;\ba) = -\kB T \log \left [\frac{\mathcal{P}_\alpha(h;\ba)}{\mathcal{P}_\alpha ( \tilde{h}_\ba^{(\alpha)};\ba )}\right ],
\end{equation}
where $\tilde{h}_\ba^{(\alpha)}$ is the most likely value of $h_\ba$ over the substrate at position $\ba$.  Therefore, $A_\alpha(h;\ba)$ is the free energy required to move the interface at equilibrium from its most likely position $\tilde{h}_\ba$ to a height $h$.  Figure~\ref{fig:hflucts} shows $A_\alpha(h;\ba)$ for a surface $\alpha$ with $f=0.05$ and $d= \l$.  The weak adhesive interaction, $\Delta \gamma = 0.05 \epsilon$, causes the $h=0$ interfacial configuration to be favored by the interface.  Fluctuations into $h=\l$ are thermally accessible to the interface ($A(h;\ba)\sim \kB T$) and for many regions, configurations with $h=2\l$ (interface is $0.4$~nm from the surface) are accessible through fluctuations of less than $3 \kB T$.  Recall that for $d=3\l$ with weak adhesions turned on, the interface is on average pulled very close to the substrate (Fig.~\ref{fig:meanheight3}), and solvent volume adjacent to the substrate has a more liquid-like density (Fig.~\ref{fig:PofN}).  But the distribution $A_\alpha(h;\ba)$ shows that local, fluctuations in the interface through the second layer are within the range of thermal fluctuations.  On the other hand, the free energy to pull the interface off of a hydrophilic patch is quite large.  These features lead to the contrasting shapes of the $\mathcal{P}_\alpha(\rho)$'s shown in Fig.~\ref{fig:PofN}, as well as the response of $\langle m_1 \rangle_{\rho,\alpha}$ to $\rho$ shown in Fig. \ref{fig:rho_dist}.

\section{Implications}
Many meso-scale solutes in nature are patchy.  Hua et. al~\cite{berne1} characterized the distributions of hydrophobic and hydrophilic subunits on an assembling surface of proteins by coarse graining the hydrophobicity on protein surfaces over $5 \times 5 \mathrm{\AA}^2$ squares.  The resulting hydrophobic surface distributions look like hydrophilic sites distributed on a hydrophobic background, not unlike the surface models we are considering.  The relative fraction of hydrophilic sites depends upon the protein considered and coincide qualitatively to our model with $f \approx 0.25-0.50$ and hydrophilic patch size $d \approx 2\l-3 \l$.

The protein surfaces studied in Ref.~\cite{berne1} and~\cite{LHpc} are finite in size, extending over several nanometers.  Our model surfaces are effectively infinite and thus lack a boundary term which may be significant in the actual kinetics of assembly.  Nonetheless, for solutes with hydrophobic surfaces extending over about a 1~nm, the mechanism for assembly can involve the drying of the solvent volume between the two assembling proteins~\cite{AL00,XH05,AW08}.  This drying event and subsequent solute aggregation is preempted by the formation of a vapor tunnel which arises when the interfaces surrounding both solutes come close to contact.  The height fluctuations of the interface next to these solutes describe this phenomenon.  Specifically, these interfacial fluctuations set the range over which solutes can broadcast their presence into the bulk solvent.  Figure~\ref{fig:hflucts2} shows the free energy $A_\alpha(h;\ba)$ for a substrate $\alpha$ with $f = 0.25$ and $d=3\l$ chosen to resemble the assembling surfaces of the proteins in Ref.~\cite{berne1}.  For surfaces consistent with these parameters, the typical range which is thermally accessible to fluctuations of the interface height is $0-2\l$ ($0.0-0.4$~nm), implying that two solutes with similar surface distributions in water can form a vapor tunnel at separations smaller than about 0.8~nm.  Not suprisingly, therefore, Hua et. al~\cite{berne1} using atomistic simuations with patchy hydrophobic protein pairs find spontaneous dewetting at surface separations of 0.6nm or less in a reasonably short period of time (within 100~ps).

\begin{figure}[ht]
  \centering
\includegraphics{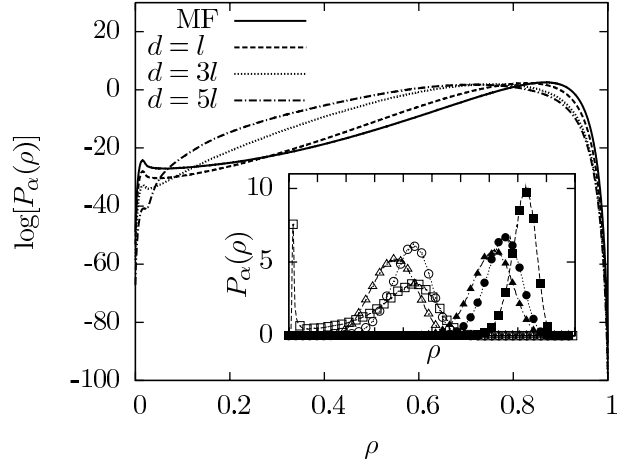}
  \caption{The probability distribution for density within the first two layers of lattice sites with weak adhesive interactions ($\Delta \gamma = 0.05 \epsilon$) over the substrate surface for specific realizations of patchy substrates with hydrophilic fraction $f = 0.03$ and patch sizes $d = \l$, $3\l$, and $5\l$. The curve label ``MF'' is the result for the mean field distribution, where the substrate is uniformly  attracted to the solvent with an attractive strength equal to $f  \epsilon$~\cite{MFE}.  The inset compares the distributions for $d=\l$ (squares), $d=3\l$ (circles), and $d = 5\l$ (triangles) for substrates with (hollow symbols) and without (filled symbols) weak adhesive interactions. }
  \label{fig:PofN}
\end{figure}

\begin{figure}[ht]
  \centering
\includegraphics{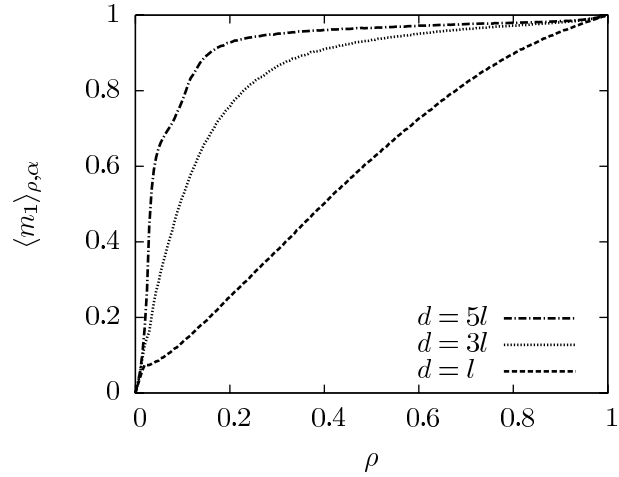}
  \caption{The mean solvent density over the hydrophilic patches, $\langle m_1 \rangle_{\rho,\alpha}$ (see text for definition), as a function of the total solvent density in the first two solvent layers, $\rho$.  Each curve is averaged over a fixed substrate realization, $\alpha$, with $f=0.03$ and weak surface adhesive interactions ($\Delta \gamma = 0.05 \epsilon$).}
  \label{fig:rho_dist}
\end{figure}

\begin{figure}[ht]
\includegraphics{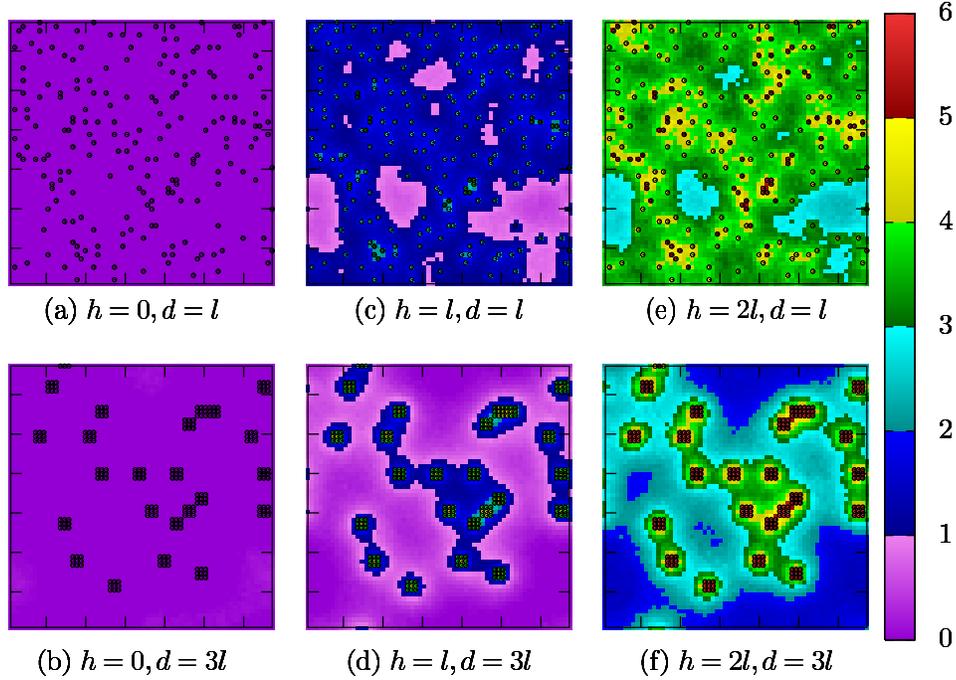}
\caption{The spatial variation of the free energy, $\beta A_\alpha(h;\ba)$ to displace the interafce at $\ba$ from its most likely height to the indicated height $h$ for a surface realization with $f = 0.05$ and with $d = \l$ and $3\l$ in the presence of weak adhesive interactions ($\Delta \gamma = 0.05 \epsilon$).  The surfaces pictured are $64\times 64~\l^2$ in size.}
  \label{fig:hflucts}
\end{figure}

\begin{figure}[ht]
  \centering
  \includegraphics{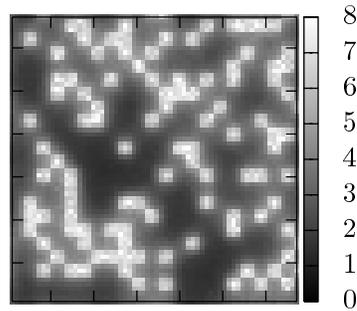}
  \caption{The spatial variation of the free energy $\beta A_\alpha(h;\ba)$ for $h = 2\l$, and a patch pattern realization with $f = 0.25$ and $d=3\l$.  The pattern of the patchy substrate $\lbrace \sigma_\ba \rbrace$ is not explicitly displayed but is evident in the pattern of $\beta A_\alpha(h;\ba)$.  The surface pictured is $64\times 64~\l^2$ in size.}
  \label{fig:hflucts2}
\end{figure}

The expulsion of water over buried hydrophilic patches presents a barrier to the water mediated assembly of patchy solutes.  For an isolated patchy surface, large hydrophilic patches remain ``wet'' until the final stages of drying.  This is because drying these large patches is very costly, as manifest in the downward curving tails for $d=5$ in Fig.~\ref{fig:PofN} near $\rho=0$.  These downward tails do not exist for $f=0$ (not shown), which is consistent with the finding of Ref.~\cite{NG07} that the addition of a hydrophilic site to an otherwise hydrophobic surface significantly slows drying.  The free energy associated with fluctuations of the interface to $h=2\l$ (0.41nm) over the hydrophilic patches are close to $10 \kB T$ (Fig.~\ref{fig:hflucts2}).  With free energy barriers this large it is unlikely that the first stage of assembly occurs through a complete drying of the assembling surfaces.  Thus, during assembly, water molecules over large hydrophilic regions are likely expelled in the latter stages of assembly through a mechanism different than those which are responsible for drying the hydrophobic regions.  In fact, retaining some amount of buried water until the final stages of assembly can be advantageous.  Specifically, water retained between assembling protein surfaces can aid in the final stages of  successful aggregation~\cite{YL04,CB98}.  Indeed, the grouping of hydrophilic regions can allow for much larger fluctuations of the interface over the substrate.  This is demonstrated in Fig.~\ref{fig:PofN} through the comparison of the distribution $\mathcal{P}_\alpha(\rho)$ for homogeneous and patchy surfaces.  Along with results for patchy substrates, Fig.~\ref{fig:PofN} has density distributions for systems with substrate layers of uniform attractive interactions with magnitude $f$ (these are the ``mean field'' lines in Fig.~\ref{fig:PofN})~\cite{MFE}.  Large fluctuations in solvent density towards the vapor-like state are less likely in the case of the homogenous substrate.  We conclude that as geometry of a patchy hydrophobic surface considerably affects the mean behavior of the adjacent solvent, it also significantly affects the interfacial fluctuations and thus can strongly influences the kinetics of assembly.

\section*{Acknowledgments}
This work was supported in its initial stages by the Director, Office of Science, Office of Basic Energy Sciences, Chemical Sciences, Geosciences, and Biosciences Division, U.S. Department of Energy under Contract No. DE-AC02-05CH11231, and then by the National Institutes of Health.  We are grateful to Lan Hua, Bruce Berne, and their co-workers for sharing with us their unpublished data~\cite{LHpc} relating to the surface distributions of self-assembling proteins.

%The references should start on their own page.

%Please compile a list of all figure captions on a separate page:

\clearpage

\begin{list}{}{\leftmargin 2cm \labelwidth 1.5cm \labelsep 0.5cm}

\item[\bf Fig. 1] The equilibrium height, $\langle h \rangle_{f,d}$, of the liquid vapor interface over the surface layer as a function of patch size $d$ for the lattice gas model (top) and the fluid interface model (bottom) at different hydrophilic fractions $f$.  In each panel, the top three of curves are computed in the absence of weak water-substrate adhesive interactions, i.e., $\Delta \gamma = \Delta \Gamma = 0$, and the bottom three curves are computed with weak water-substrate adhesive interactions.

\item[\bf Fig. 2] Mean interfacial height, $\mhxy$ projected onto the $xy-$plane for hydrophilic fraction $f = 0.03$ and $d = \l$ (top left), $d=2 \l$ (top right), and $d=3 \l$ (bottom left and right).  The height $h$ is indicated by the shading (scale at right)and the location of hydrophilic surface sites (substrate sites with $\sigma_\ba = 1$) are shown with red circles.  Panels (a), (b), and (c) correspond to averages in the absence of surface adhesive interactions  ($\Delta \gamma = 0.0$) and panel (d) corresponds to an average with surface adhesive interactions ($\Delta \gamma = 0.05 \epsilon$).  The surfaces pictured are $64\times64 \; \l^2$ in size.

\item[\bf Fig. 3] The probability distribution for density within the first two layers of lattice sites with weak adhesive interactions ($\Delta \gamma = 0.05 \epsilon$) over the substrate surface for specific realizations of patchy substrates with hydrophilic fraction $f = 0.03$ and patch sizes $d = \l$, $3\l$, and $5\l$. The curve label ``MF'' is the result for the mean field distribution, where the substrate is uniformly  attracted to the solvent with an attractive strength equal to $f  \epsilon$~\cite{MFE}.  The inset compares the distributions for $d=\l$ (squares), $d=3\l$ (circles), and $d = 5\l$ (triangles) for substrates with (hollow symbols) and without (filled symbols) weak adhesive interactions.

\item[\bf Fig. 4] The mean solvent density over the hydrophilic patches, $\langle m_1 \rangle_{\rho,\alpha}$ (see text for definition), as a function of the total solvent density in the first two solvent layers, $\rho$.  Each curve is averaged over a fixed substrate realization, $\alpha$, with $f=0.03$ and weak surface adhesive interactions ($\Delta \gamma = 0.05 \epsilon$).

\item[\bf Fig. 5] The spatial variation of the free energy, $\beta A_\alpha(h;\ba)$ to displace the interafce at $\ba$ from its most likely height to the indicated height $h$ for a surface realization with $f = 0.05$ and with $d = \l$ and $3\l$ in the presence of weak adhesive interactions ($\Delta \gamma = 0.05 \epsilon$).  The surfaces pictured are $64\times 64~\l^2$ in size.

\item[\bf Fig. 6] The spatial variation of the free energy $\beta A_\alpha(h;\ba)$ for $h = 2\l$, and a patch pattern realization with $f = 0.25$ and $d=3\l$.  The pattern of the patchy substrate $\lbrace \sigma_\ba \rbrace$ is not explicitly displayed but is evident in the pattern of $\beta A_\alpha(h;\ba)$.  The surface pictured is $64\times 64~\l^2$ in size.

\end{list}

\clearpage

\end{document}